\documentclass[aps,prl,reprint,nofootinbib,nobibnotes,amsmath,amssymb,superscriptaddress,floatfix]{revtex4-2}
\usepackage[T1]{fontenc}
\usepackage[utf8]{inputenc}
\usepackage[english]{babel}
\usepackage{graphicx}
\usepackage{mathtools}
\usepackage{bm}
\usepackage[dvipsnames]{xcolor}
\IfFileExists{orcidlink.sty}{\usepackage{orcidlink}}{\newcommand{\orcidlink}[1]{}}
\usepackage{ulem}
\usepackage{hyperref}
\definecolor{mypurple}{RGB}{128, 0, 128}
\usepackage{hyperref}
\hypersetup{
    colorlinks=true,
    linkcolor=mypurple, 
    citecolor=mypurple, 
    urlcolor=mypurple  
 }
\newcommand{\newsec}[1]{\textit{\textbf{#1}}}
\newcommand{\ed}{\mathop{}\!\mathrm d}
\newcommand{\ab}[1]{\left|#1\right|}
\newcommand{\br}[1]{\left[#1\right]}
\newcommand{\pa}[1]{\left(#1\right)}

\newcommand{\MaxPlanck}{\affiliation{Max Planck Institute for
Gravitational
      Physics (Albert Einstein Institute), Am M{\"u}hlenberg 1, D-14476
Potsdam,
      Germany}}
\newcommand{\Cornell}{\affiliation{Cornell Center for Astrophysics and Planetary Science, Cornell University, Ithaca, New York 14853, USA}}
\newcommand{\Caltech}{\affiliation{Theoretical Astrophysics, Walter Burke
  Institute for Theoretical Physics, California Institute of Technology,
  Pasadena, California 91125, USA}}

\begin{document}

\title{Quadratic Gravitational-Wave Scattering by Kerr Black Holes}

\author{Lennox S. Keeble\,\orcidlink{0009-0009-5796-631X}}
\email{keebls25@wfu.edu}
\affiliation{Department of Physics, Wake Forest University, Winston-Salem, North Carolina 27109, USA}
\author{Hengrui Zhu \orcidlink{0000-0001-9027-4184}}
\email{hengruizhu0330@gmail.com}
\affiliation{Department of Physics, Princeton University, Princeton, New Jersey 08544, USA}
\affiliation{Leinweber Forum for Theoretical Physics, Princeton University, Princeton, New Jersey 08544, USA}
\author{Lawrence E.~Kidder \orcidlink{0000-0001-5392-7342}} \Cornell
\author{Harald P.~Pfeiffer \orcidlink{0000-0001-9288-519X}} \MaxPlanck
\author{Mark A.~Scheel \orcidlink{0000-0001-6656-9134}} \Caltech
\date{\today}

\begin{abstract}
Second-order black-hole perturbation theory will be an important component of precision gravitational-wave modeling and tests of strong-field gravity with next-generation detectors.
Beyond quasinormal modes (QNMs), whose quadratic interactions have been the focus of recent work in black-hole spectroscopy, a generic retarded solution consists of a continuum of real-frequency scattering states whose nonlinear interactions in Kerr remain comparatively less explored.
We use numerical relativity to study the quadratic response of Kerr black holes to  nearly monochromatic $(\ell,m)=(2,\pm2)$ incident gravitational waves, measuring the complex self-coupling to the outgoing $(\ell,m)=(4,4)$ daughter at twice the parent frequency for black-hole spins up to $a=0.95$.
At low frequencies, the coupling is strongly suppressed and exhibits opposite spin dependence for prograde and retrograde scattering. 
At higher prograde frequencies, we identify a daughter-mode resonance whose frequency and width extracted from the nonlinear response track the fundamental $(\ell,m,n)=(4,4,0)$ QNM. Near the fundamental $(2,2,0)$ QNM frequency, the in-mode coupling grows with spin, in contrast to the decreasing quadratic QNM coupling, demonstrating that the nonlinear response depends on the full parent scattering state rather than on its frequency alone.
A complementary semi-analytic second-order Teukolsky calculation reproduces the nonlinear response from our numerical relativity simulations. Our numerical-relativity results and their semi-analytic extension provide a basis for generic homogeneous radiative perturbations of Kerr at second order, with applications to dynamical tides, near-extremal dynamics, and second-order gravitational self-force theory.
\end{abstract}
\maketitle

\newsec{Introduction.}
The response of a Kerr black hole to gravitational perturbations is central to a range of strong-field problems, including black-hole ringdown, dynamical tides, and extreme-mass-ratio inspirals. At linear order, these phenomena are described by the Teukolsky equation~\cite{Teukolsky:1972my,Teukolsky:1973ha,Teukolsky:1974yv,Sasaki:2003xr}. For each separated angular mode and generic real frequency, the homogeneous radial solution space is spanned by the ``in'' and ``up'' solutions~\cite{Teukolsky:1974yv,Mano:1996vt,Sasaki:2003xr}. The in-mode describes a wave incident from past null infinity, partly reflected to future null infinity and partly transmitted through the future horizon, while the up-mode is the complementary solution originating from the past horizon.

The same pair of homogeneous solutions is used, through their Wronskian, to construct the radial Green's function for sourced perturbations. Upon analytic continuation in frequency, this Green's function exhibits QNM poles together with non-pole contributions associated with prompt response and late-time tails~\cite{Price:1971fb,Leaver:1986gd,Green:2022htq}. The in- and up-mode solutions therefore underlie both frequency-domain self-force calculations~\cite{Hughes:2021exa,Barack:2018yvs,Pound:2021qin} and scattering-based descriptions of compact-object tidal response~\cite{Bautista:2021wfy,Bautista:2022wjf,Ivanov:2024sds,Saketh:2023bul}. QNMs capture an important resonant part of this response, but not the full linear dynamics. Understanding how the underlying scattering solutions couple is therefore a natural step towards characterizing the generic second-order response of Kerr.

At second order, the radiative perturbation obeys~\cite{Campanelli:1998jv}
\begin{equation}
 \mathcal T\bigl[\psi_4^{(2)}\bigr]
 =\mathcal S^{(2)}\bigl[h^{(1)},\psi_4^{(1)}\bigr],
 \label{eq:second_order_intro}
\end{equation}
where $h^{(1)}_{ab}$ is the first-order metric perturbation, $\psi_4^{(n)}$ is the $n$th-order Weyl scalar, and $\mathcal T$ is the linear Teukolsky operator. The source is quadratic in the first-order fields and their derivatives. Expanding a homogeneous linear perturbation in the scattering basis therefore expresses its quadratic source as a sum of pairwise couplings. Hence, the in-in, in-up, and up-up couplings organize the second-order radiative solution. A complete second-order solution requires summing these couplings over all parent frequencies, angular channels, and polarizations, and supplementing the radiative sector with the non-radiative metric perturbations and any independent homogeneous second-order solution~\cite{Green:2019nam,Spiers:2023cip}.

Second-order perturbation theory has been developed extensively in Schwarzschild~\cite{Gleiser:1995gx,Ioka:2007ak,Nakano:2007cj,Brizuela:2006ne,Brizuela:2009qd,Pazos:2010xf,Spiers:2023mor} and Kerr spacetimes~\cite{Campanelli:1998jv,Green:2019nam,Loutrel:2020wbw,Ripley:2020xby,Spiers:2023cip}. Much of the recent work has concerned quadratic QNMs (QQNMs), whose frequencies follow from products of QNM parent fields and their complex conjugates. These modes have been identified in numerical merger waveforms and studied in different scenarios using numerical relativity (NR) and perturbation theory~\cite{Papadopoulos:2001zf,Zlochower:2003yh,London:2014cma,Cheung:2022rbm,Mitman:2022qdl,Kehagias:2023ctr,Redondo-Yuste:2023seq,Perrone:2023jzq,Zhu:2024rej,Cheung:2023vki,Ma:2024qcv,Bucciotti:2024zyp,Bucciotti:2024jrv,Bourg:2024jme,Khera:2024bjs,Lagos:2024ekd,Bucciotti:2025rxa,Bourg:2025lpd}. The corresponding couplings between real-frequency scattering states have been much less explored~\cite{Ma:2025rnv,Cardoso:2026llh, Bucciotti:2026quad}.

\begin{figure}[!t]
 \centering
 \includegraphics[width=\columnwidth]{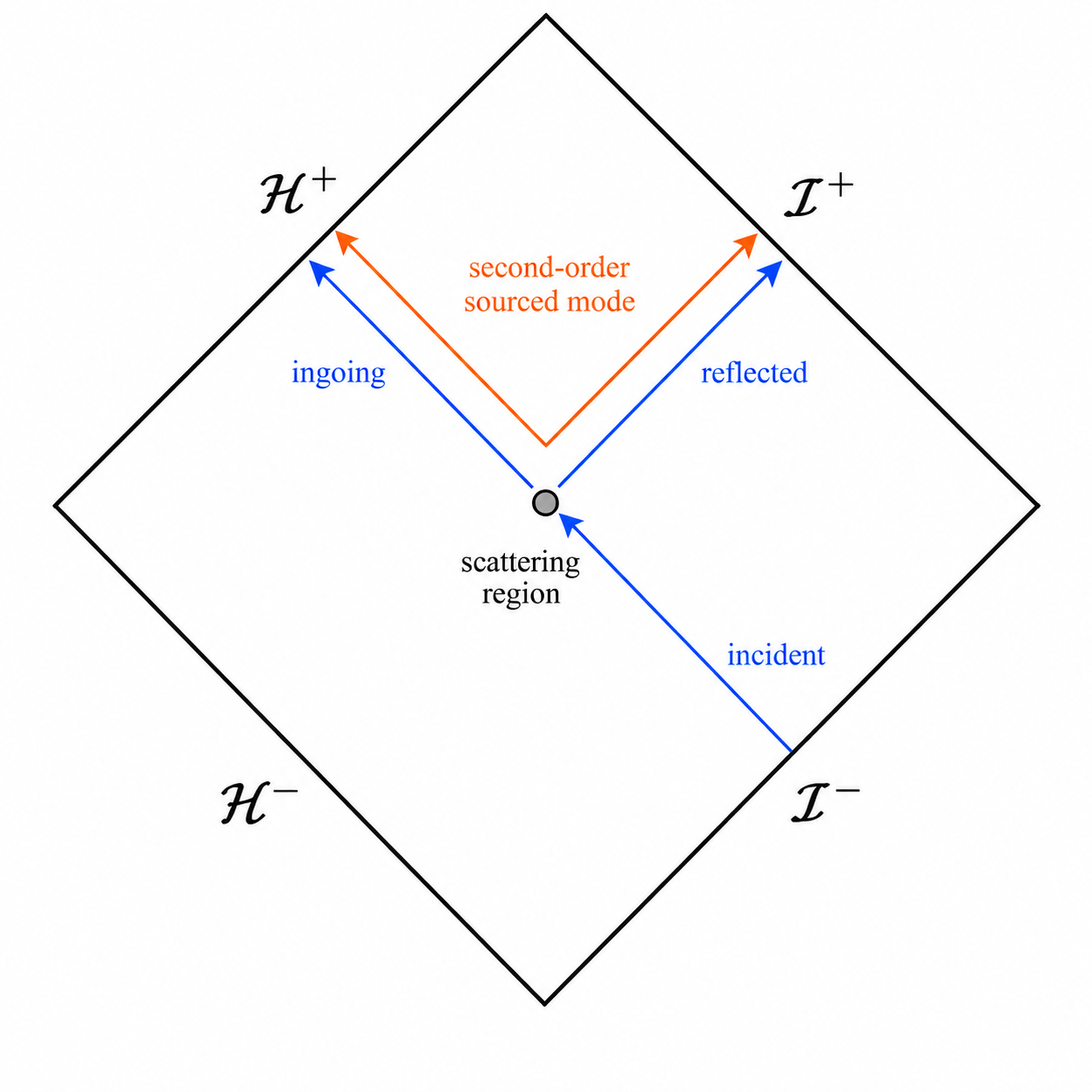}
 \caption{Schematic Penrose diagram of quadratic wave generation. Blue arrows show a linear in-mode incident from past null infinity $\mathcal{I}^{-}$ and its reflected and transmitted components. Orange arrows indicate second-order radiation propagating to future null infinity $\mathcal I^+$ and through the future horizon $\mathcal H^+$. The interaction is distributed throughout the exterior rather than confined to the point shown.}
 \label{fig:penrose_diag}
\end{figure}

As illustrated in Fig.~\ref{fig:penrose_diag}, a linear in-mode sources second-order radiation propagating both towards the horizon and to infinity, with the dynamics governed by Eq.~\eqref{eq:second_order_intro}. The work of Ref.~\cite{Cardoso:2026llh} recently studied higher-harmonic generation from axisymmetric wave scattering in Schwarzschild, finding daughter-mode resonances and low-frequency suppression; Ref.~\cite{Bucciotti:2026quad} has extended this program in Schwarzschild to more generic angular structure and parity. Rotation introduces additional structure: frame dragging distinguishes prograde and retrograde waves, absorption changes sign across the superradiant frequency, and both the radial propagation and QNM spectrum depend on spin~\cite{Teukolsky:1974yv,East:2013mfa,Brito:2015oca}. Long-lived modes near extremality make resonant nonlinear interactions particularly relevant~\cite{Detweiler:1980gk,Yang:2012pj,Yang:2013uba,Yang:2014tla,Gralla:2016sxp,Iuliano:2024ogr,Lehner:2026tfe}.

In this work, we measure the equal-frequency in-mode self-coupling
$(\ell,\pm m,\omega)=(2,\pm2,\omega)$ into the daughter channel
$(4,4,2\omega)$, which we denote
$(2,\pm2,\omega)^2\to(4,4,2\omega)$, using fully nonlinear NR scattering experiments. We send long quadrupolar wave trains towards Kerr black holes with dimensionless spins $0\leq a\leq0.95$ and extract the outgoing quadratically generated daughter radiation. 
Varying the parent frequency gives a complex coupling spectrum, which we define as the outgoing daughter-mode strain amplitude normalized by the square of the parent's.

We find low-frequency coupling suppression and, in this regime, opposite spin dependence for prograde and retrograde coupling. At higher prograde frequencies, resonant amplitude and phase features track the fundamental daughter $(\ell,m,n)=(4,4,0)$ QNM, where $n$ labels the overtone index with increasing decay rate. We also find that the in-mode coupling grows with spin for parent frequencies $\omega=\mathrm{Re}[\omega_{220}(a)]$, in contrast to the corresponding QQNM coupling which is suppressed with spin~\cite{Zhu:2024rej,Redondo-Yuste:2023seq,Ma:2024qcv,Khera:2024bjs}. The measured complex response also agrees closely with a complementary semi-analytic second-order Teukolsky calculation, whose derivation and implementation will be presented separately~\cite{ZhuInPrep}. We work in units in which $G=c=1$ and define $a=J/M^2$. Unless shown explicitly, we set the initial black-hole mass $M=1$. Positive and negative $\omega$ label prograde and retrograde incident $m=2$ waves.

\begin{figure*}[!t]
 \centering
 \includegraphics[width=\textwidth]{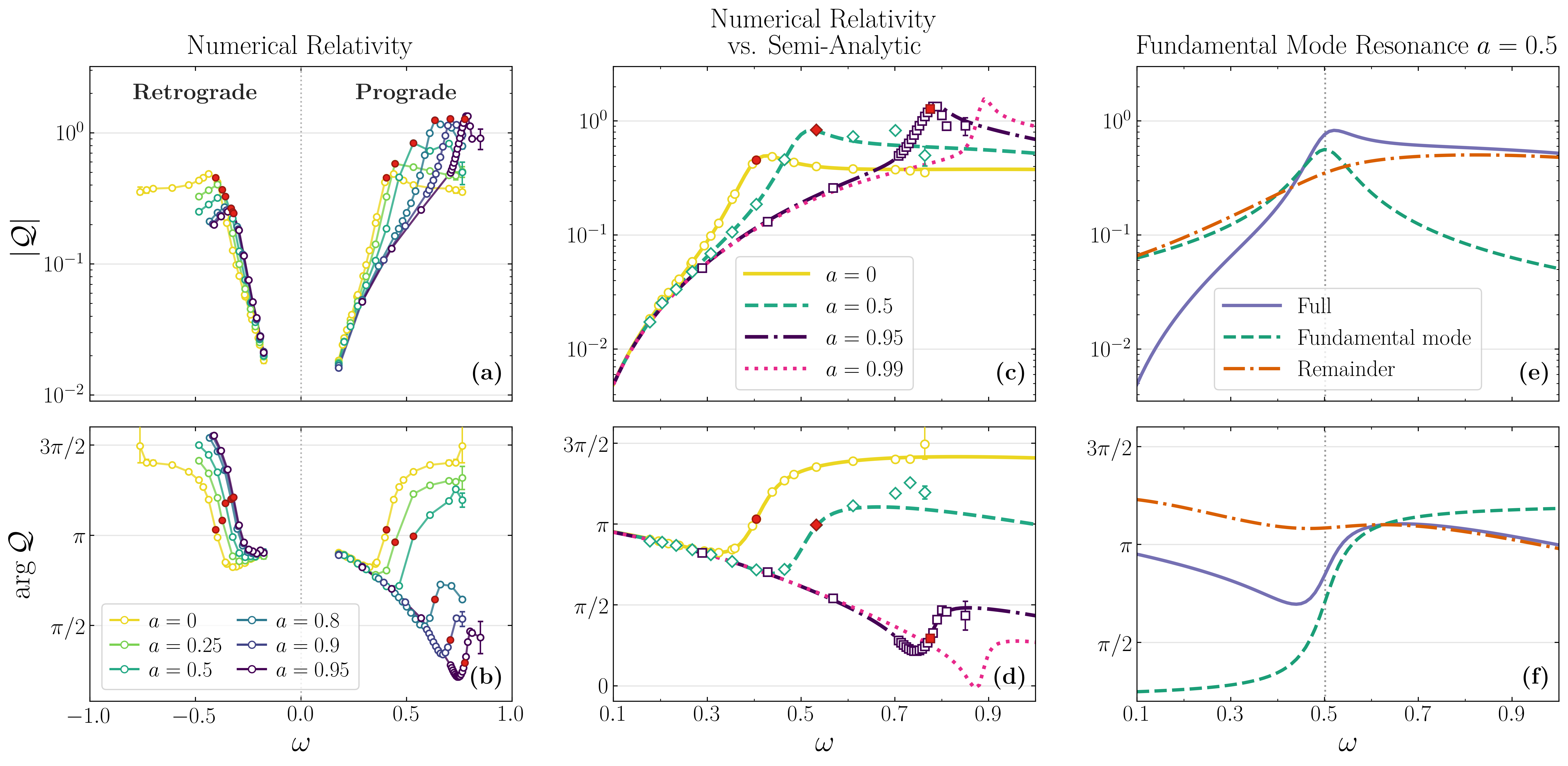}
 \caption{\textbf{Quadratic in-mode coupling spectrum.} Upper and lower rows show the modulus and phase, unwrapped separately on each branch, of the quadratic coupling $\mathcal{Q}$ as a function of the parent frequency $\omega$ for various fixed spins indicated in the legends. (a,b) Numerical relativity spectrum. Red markers denote frequency samples nearest the daughter resonance $\mathrm{Re}(\omega_{440})/2$. (c,d) Circles, diamonds, and squares denote NR coupling at $a\in\{0,0.5,0.95\}$; the curves correspond to semi-analytic results from Ref.~\cite{ZhuInPrep}, with an additional curve for $a=0.99$. (e,f) Semi-analytic response at $a=0.5$ decomposed into its fundamental $(4,4,0)$ pole contribution and the complex remainder; the vertical lines mark $\mathrm{Re}(\omega_{440})/2$. The NR and semi-analytic results agree closely in both phase and amplitude, showing more deviation at higher frequencies where (i) the NR data become less accurate due to the diminishing reflected amplitude and (ii) spherical-spheroidal mixing grows (the semi-analytic calculation uses spheroidal modes whereas NR uses spherical extraction), the effect of which saturates at less than $2$\% for the spins and frequencies considered here. The NR error bars account for fit and extrapolation sensitivity.}
 \label{fig:NRcoupling}
\end{figure*}

\begin{figure}[!t]
 \centering
 \includegraphics[width=\columnwidth]{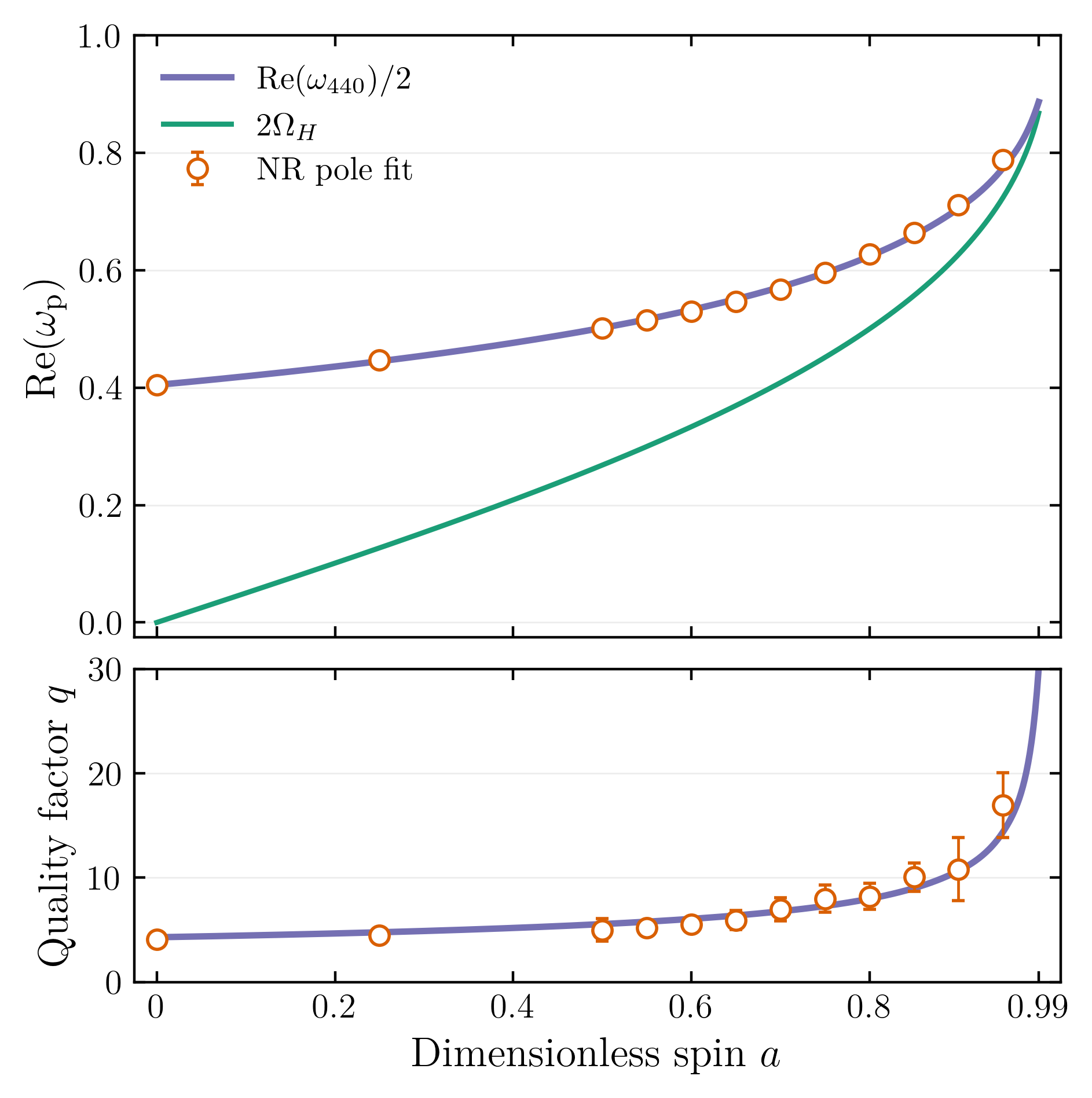}
 \caption{\textbf{Daughter resonance with spin.} NR estimates of the complex pole give its real frequency (orange circles, upper panel) and quality factor $q=\mathrm{Re}(\omega_{\rm p})/[2|\mathrm{Im}(\omega_{\rm p})|]$ (lower panel). The purple curve is computed directly from the fundamental QNM frequency $\omega_{\rm p}=\omega_{440}(a)/2$~\cite{Leaver:1985ax,Stein:2019mop}. The green curve, drawn for reference, is the superradiant frequency $2\Omega_H\equiv 2a/(r_+^2+a^2)$, where $r_{+}$ is the outer horizon radius. The superradiant frequency converges towards the QNM frequencies at extremality, coinciding with the sharpening resonant response, as indicated by the diverging quality factor.}
 \label{fig:resonance_tracking}
\end{figure}

\begin{figure}[!t]
 \centering
 \includegraphics[width=\columnwidth]{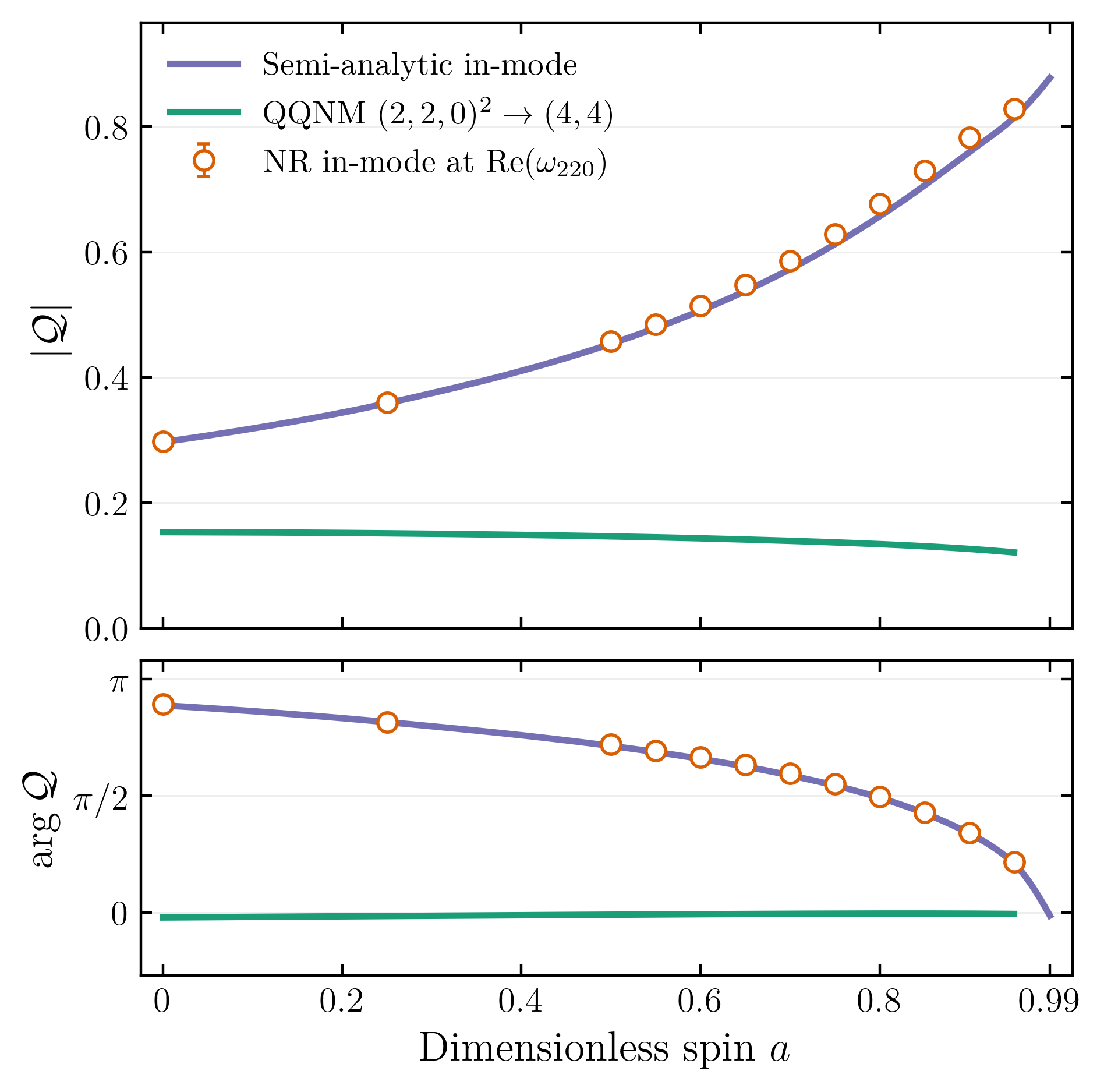}
 \caption{\textbf{In-mode versus QQNM coupling.} Orange circles denote coupling magnitude and phase (upper and lower panels, respectively) from NR simulations with parent frequencies $\omega=\mathrm{Re}[\omega_{220}(a)]$ (the $a\in\{0,0.25\}$ cases require frequency interpolation). The green curve is obtained from the $(2,2,0)^2$ QQNM fit published in Ref.~\cite{Khera:2024bjs}. The purple curve is a complementary semi-analytic in-mode calculation~\cite{ZhuInPrep}. Error bars include fit, extrapolation, and, where needed, interpolation sensitivity. The in-mode magnitude increases with spin, in contrast to the QQNM magnitude which decreases.}
 \label{fig:qnm_coupling_comparison}
\end{figure}

\newsec{Numerical setup.}
We use the Spectral Einstein Code (\textsc{SpEC}) to evolve the first-order generalized-harmonic system on a multi-domain pseudo-spectral grid with black-hole excision~\cite{Lindblom:2005qh,Scheel:2006gg}. Following Refs.~\cite{Zhu:2024rej,Zhu:2024dyl}, we perturb a spherical Kerr--Schild black hole~\cite{Chen:2021rtb} with a finite, rotating electric-parity quadrupolar wave train. The conformal-metric perturbation contains the $m=\pm2$ partners of a normalized tensor harmonic, with a smooth radial envelope and an approximately constant rescaled amplitude on its plateau. We prescribe an ingoing time derivative, remove its conformal trace, and solve the extended conformal thin-sandwich equations~\cite{Pfeiffer:2002iy,Pfeiffer:2004qz}. During the evolution, the generalized-harmonic gauge source is held fixed at its initial value to minimize unwanted gauge dynamics. The perturbation amplitude is chosen to resolve the daughter signal while limiting changes in the horizon mass and spin to less than $1\%$. Further details pertaining to the initial data are described in the End Matter.

We extract $r\psi_4^{\ell m}$ in spin-weighted spherical harmonics at several radii. Complex Fourier fits over a quasi-stationary interval isolate the reflected parent at $\omega$ and the daughter at $2\omega$. We define the dimensionless quadratic coupling coefficient in the strain as
\begin{equation}
 \mathcal Q(\omega)=\frac{(M\omega)^2}{4}
 \frac{\mathcal A^{\psi_4}_{44}(2\omega)}
      {[\mathcal A^{\psi_4}_{22}(\omega)]^2},
 \label{eq:strain_coupling}
\end{equation}
where $\mathcal A^{\psi_4}_{\ell m}$ denotes the dimensionless amplitude of $Mr\psi_4^{\ell m}$. With $\psi_4=-\partial_t^2(h_+-ih_\times)$, Eq.~\eqref{eq:strain_coupling} is the daughter strain amplitude of $(r/M)h$ divided by the complex square of the reflected parent strain amplitude. One could also define an incident-normalized coefficient $\mathcal Q\mathcal R_{22}^2$, where $\mathcal R_{22}$ is the complex linear reflection coefficient. We extrapolate the quadratic coupling coefficient $\mathcal{Q}(\omega)$ to infinity from measurements at finite radii~\cite{Boyle:2009vi}. We quote uncertainties in the extrapolated coupling by varying the temporal fit and radial extrapolation order at fixed grid resolution. A detailed convergence study near a high-spin resonance is provided in the Supplemental Material.

\newsec{Results.}
Panels (a,b) of Fig.~\ref{fig:NRcoupling} show the measured quadratic coupling spectrum from NR simulations.
The magnitude of the coupling decreases as $|\omega|\to0$ and the spin dependence weakens as the longer-wavelength incident waves probe larger-scale structures of the spacetime.
At fixed low frequency, increasing spin suppresses the prograde coupling and enhances the retrograde coupling. 
Through a phenomenological model described in the Supplemental Material, we interpret this asymmetry as arising from the spin dependence of the radial turning points: frame dragging shifts the effective potential barrier and changes the region over which the incident wave is reflected and attenuated. 
In the retrograde case, the wave penetrates more deeply into the strong field with increasing spin and is thereby concentrated over a smaller radial region, increasing the local overlap that sources the quadratic response, whereas the reverse behavior occurs for prograde waves. 

At higher prograde frequencies, the coupling develops a pronounced
resonant feature in both amplitude and phase. The resonant parent frequency closely tracks half the fundamental $(\ell,m,n)=(4,4,0)$ QNM frequency with spin, i.e., $\omega_{\mathrm{res}}\simeq \mathrm{Re}(\omega_{440})/2$. To quantify these resonances, we fit the complex quadratic coefficient to
\begin{equation}
 \mathcal Q(\omega)=\frac{C}{\omega-\omega_{\rm p}}
                   +B_0+B_1(\omega-\omega_0),
 \label{eq:pole_fit}
\end{equation}
where $C$, $B_0$, and $B_1$ are complex, $\omega_0=\mathrm{Re}(\omega_{440})/2$ is a fixed reference
frequency at which the fit window is centered, and the complex pole $\omega_{\rm p}$ is fitted freely.
As shown in Fig.~\ref{fig:resonance_tracking}, the fitted pole frequency $\omega_{\rm{p}}$ and its associated quality factor $q=\mathrm{Re}(\omega_{\rm{p}})/\left[2\ab{\mathrm{Im}(\omega_{\rm{p}})}\right]$ from the NR data closely track the corresponding values obtained from the fundamental $(4,4,0)$ QNM. Near resonance, the system thus behaves analogously to a driven damped harmonic oscillator: the real part
of $\omega_{\rm p}$ determines the resonance frequency while
$|\mathrm{Im}(\omega_{\rm p})|$ sets the resonance width and
the damping rate.

Figure~\ref{fig:qnm_coupling_comparison} compares the in-mode coupling
for parent frequencies $\omega=\mathrm{Re}[\omega_{220}(a)]$ with
$(2,2,0)^2$ QQNM coupling obtained from Ref.~\cite{Khera:2024bjs}. Despite sharing the
same real parent frequency, the two responses differ markedly in both
magnitude and phase. In particular, the in-mode coupling is larger and increases with spin, in contrast to the decreasing QQNM coupling. The difference arises because
the scattering state and QNM have distinct radial profiles and boundary
conditions, illustrating that the quadratic response depends on the details of its full parent solution rather than on its frequency alone.

Panels (c,d) of Fig.~\ref{fig:NRcoupling} compare
the NR measurements with the second-order Teukolsky calculation of
Ref.~\cite{ZhuInPrep}. The two agree closely in both amplitude and phase
across the resolved frequency range, including the resonant structure.
Along the sequence shown in Fig.~\ref{fig:qnm_coupling_comparison}, the
directly sampled calculations differ from NR by only $0.3$--$3.1\%$ in
magnitude and $0.24$--$1.32$ degrees in phase. These differences are not a complete error estimate since the two calculations
use different angular bases, i.e., the NR results do not take into account spherical-spheroidal mixing~\cite{Berti:2014fga} in both the initial data and waveform extraction.

Panels (e,f) of Fig.~\ref{fig:NRcoupling} further show how the daughter QNM
resonance is embedded in the full scattering response. We decompose the
semi-analytic result into the fundamental $(4,4,0)$ pole contribution
and the remaining response, which contains the other poles and non-pole
terms. Although the fundamental pole controls the resonant variation,
the remainder remains significant and interferes coherently with it,
shifting the amplitude maximum and modifying the phase evolution.
Thus, the daughter QNM governs the resonant structure but does not by
itself describe the full nonlinear scattering response.

\newsec{Discussion.}
Our NR experiments resolve a frequency-dependent component of Kerr's
quadratic gravitational response. We find low-frequency coupling suppression due to the increasing wavelength of the linear waves,
opposite spin dependence for prograde and retrograde scattering, and
resonant behavior associated with the fundamental daughter QNM. Along the
sequence $\omega=\mathrm{Re}[\omega_{220}(a)]$, the in-mode coupling
also exhibits a spin dependence that qualitatively differs from that of the corresponding QQNM coupling. This demonstrates that the quadratic response
depends on the full structure of the parent perturbation, rather than on
its oscillation frequency alone.

More generally, the in-in, in-up, and up-up couplings provide the
building blocks for the quadratic radiative response to generic
homogeneous linear perturbations. The results presented here focus on only one
equal-frequency channel, but extending this program to additional
frequencies and angular sectors will allow a broader characterization of
the second-order response of Kerr, as has been done recently in Schwarzschild~\cite{Bucciotti:2026quad}. Such generalizations in Kerr, which will be addressed in the semi-analytic work~\cite{ZhuInPrep}, are necessary for the following applications of our work.

One application is to computing the nonlinear dynamical tidal response of black holes.
For a quasi-circular binary, the tidal field is concentrated at discrete
harmonics of the orbital frequency, and the channel studied here directly
probes the quadratic self-coupling of one such harmonic into its second
harmonic. More general tidal configurations involve couplings among
different orbital harmonics and angular modes. Extending the present
calculation to these channels, together with an explicit matching between
the scattering amplitudes and worldline response functions~\cite{Bautista:2021wfy,Bautista:2022wjf,Ivanov:2024sds,Saketh:2023bul,Ivanov:2026icp,Ivanov:2022qqt,Chakraborty:2025wvs,Charalambous:2021mea,Combaluzier--Szteinsznaider:2025eoc,Chakraborty:2026dox}, would provide
a route towards determining nonlinear dynamical tidal response coefficients~\cite{Pani:2025qxs,DeLuca:2023mio,Riva:2023rcm,Iteanu:2024dvx,Gounis:2024hcm}.

Near extremality, several characteristic frequencies become degenerate.
In particular, the zero-damped QNM spectrum approaches the superradiant bound, i.e., 
$\mathrm{Re}(\omega_{\ell m n})\to m\Omega_H$~\cite{Detweiler:1980gk,Yang:2012pj,Yang:2013uba}.
Linear incident perturbations near this bound are increasingly reflected rather
than absorbed, allowing a rapidly rotating black hole to retain its spin.
At second order, however, the same frequency alignment can place the
quadratically generated $(4,4)$ daughter close to a long-lived resonance,
as in the coupling studied here (see lower panel of Fig.~\ref{fig:resonance_tracking}). The resulting weak damping and sustained
quadratic and higher-order driving may allow nonlinear perturbations to accumulate over
long timescales, potentially leading to gravitational turbulence, as observed in other settings~\cite{Yang:2014tla,Yang:2015jja,Ma:2025rnv,Iuliano:2024ogr}. This growth competes with other quadratic channels,
particularly axisymmetric $m=0$ modes, which do not benefit from
superradiance and can be absorbed by the horizon. Because they carry
energy but no axial angular momentum, such absorption tends to reduce the
dimensionless spin, thereby perturbing the system slightly off resonance. The nonlinear evolution near extremality may therefore
involve a competition between long-lived resonantly enhanced non-axisymmetric modes
and spin-down through non-superradiant channels. The eventual endpoint of this competition
remains an open problem~\cite{Yang:2014tla,Gralla:2016sxp,Iuliano:2024ogr,Lehner:2026tfe}.

Lastly, these scattering experiments provide useful vacuum benchmarks
for higher-order self-force calculations. Extending the present
construction to particle-sourced perturbations~\cite{Wardell:2021fyy} additionally requires
source matching, metric reconstruction, and regularization, but the same
quadratic scattering channels enter the nonlinear radiative sector~\cite{Pound:2012nt,Barack:2018yvs,Pound:2021qin,Spiers:2023cip,Bourg:2024vre,Leather:2026zhl}.

\newsec{Acknowledgments.}
We thank Alejandro C\'ardenas-Avenda\~no, Abhishek Hegade, Luis Lehner, Sizheng Ma, Frans Pretorius, and Haiyang Wang for insightful discussions. We also thank Jaime Redondo-Yuste for facilitating a comparison between our works in the Schwarzschild setting. H.Z. especially thanks Frans Pretorius and L.K. thanks Alejandro C\'ardenas-Avenda\~no and Dorothy Cutler for repeated discussions. The simulations presented in this article were performed on computational resources managed and supported by Princeton University's Research Computing. OpenAI ChatGPT versions $5.6$ and $6$ and Anthropic Claude Fable $5$ were used to accelerate data post-processing and manuscript proofreading. All scientific analyses, numerical results, and conclusions were produced by the authors, who take full responsibility for the content of this work.

\normalem
\bibliographystyle{apsrev4-2}
\bibliography{main_refs}
\ULforem

\clearpage
\section*{End Matter: Initial data construction}
\label{sec:initial_data}
We give the conformal-metric seed and its time derivative in units $M=1$. Let $q_{AB}$ and $D_A$ be the metric and covariant derivative on the unit sphere, and normalize the scalar harmonics by $\int|Y_{\ell m}|^2\,\mathrm d\Omega=1$. The electric-parity tensor harmonic is~\cite{Martel:2005ir}
\begin{equation}
\begin{aligned}
 E^{\ell m}_{AB}
 &=\sqrt{\frac{2}{\lambda(\lambda-2)}}
   \left(D_AD_B+\frac{\lambda}{2}q_{AB}\right)Y_{\ell m},\\
 \lambda&=\ell(\ell+1),\qquad
 \int E^{\ell m}_{AB}E^{AB\,\ell m*}\,\mathrm d\Omega=1,
\end{aligned}
\label{eq:electric_harmonic}
\end{equation}
where indices are raised with $q^{AB}$.

The nonzero angular components of the freely specified perturbation are
\begin{equation}
\begin{aligned}
 h_{AB}&=\sqrt{96}\,\epsilon rW(y)
          \mathrm{Re}\!\left[E^{22}_{AB}e^{-i\eta\psi(y)}\right],\\
 y&=r-r_0+t,\qquad r_0=30,\\
 \psi(y)&=\omega_c y+\pi/4,\qquad \omega_c>0,\\
 W(y)&=\tanh y+\tanh(L-y),\qquad L=2\pi N/\omega_c.
\end{aligned}
\label{eq:initial_seed}
\end{equation}
Here, $h_{rr}=h_{rA}=0$, $N$ controls the train length, $\epsilon$ is a smallness parameter for the perturbation amplitude, and $\eta=\pm1$ sets the direction of rotation, with carrier frequency $\omega=\eta\omega_c$. Taking the real part includes the conjugate azimuthal partner required for a real spatial metric. Equivalently, the two relative orientations can be obtained by reversing the background spin at fixed positive carrier frequency.

In an orthonormal angular frame, $h_{\hat A\hat B}\propto W/r$ and
\begin{equation}
\begin{aligned}
 \partial_t h_{\hat A\hat B}
 &=\frac{\sqrt{96}\,\epsilon}{r}
 \mathrm{Re}\!\left[
   (W'-i\eta\omega_c W)E^{22}_{\hat A\hat B}e^{-i\eta\psi}
 \right],\\
 \partial_r h_{\hat A\hat B}
 &=\partial_t h_{\hat A\hat B}-h_{\hat A\hat B}/r,
\end{aligned}
\label{eq:seed_derivatives}
\end{equation}
where the explicit factor of $r$ is held fixed when taking the time derivative. Cartesian derivatives also act on the angular basis. Harmonic projection and derivative checks verify the prescribed seed before solving the constraints.

On the spherical Kerr--Schild background $\bar\gamma_{ij}$, we set
$\tilde\gamma_{ij}=\bar\gamma_{ij}+h_{ij}$ and
$\tilde u_{ij}=(\partial_t h_{ij})^{\mathrm{TF}}_{\tilde\gamma}$.
The remaining initial data are obtained by solving the extended conformal thin-sandwich equations. Angular trace freedom alone does not make the seed an exactly transverse vacuum perturbation at finite radius. The solved curvature can therefore contain quadratic $(4,4)$ and axisymmetric contributions even when the freely specified seed has only the intended quadrupolar harmonic. We retain these nonlinear contributions.

\clearpage
\section*{Supplemental Material}
\appendix
\setcounter{equation}{0}
\renewcommand{\theequation}{S\arabic{equation}}
\renewcommand{\theHequation}{supp.\arabic{equation}}
\setcounter{figure}{0}
\renewcommand{\thefigure}{S\arabic{figure}}
\renewcommand{\theHfigure}{supp.\arabic{figure}}
\setcounter{table}{0}
\renewcommand{\thetable}{S\arabic{table}}
\renewcommand{\theHtable}{supp.\arabic{table}}

\section{Numerical-relativity diagnostics}
\label{app:convergence}
We describe the resolution study, waveform fits, and extrapolation used to assess the measured complex coupling. All numerical amplitudes and coordinates in this supplement use $M=1$ unless dimensions are displayed explicitly.

\subsection{Evolution resolution}
The controlled test uses spin $a=0.95$ and frequency $M\omega=0.7752604113$, which is near the fundamental daughter resonance. Four evolution grids start from the same initial data, constructed on 33 subdomains with radial order $N_r=18$ and angular truncation $N_\ell=22$, and evolve to $t=300M$. Each evolution grid contains 56 subdomains. Level $k$, with $k\in\{0,1,2,3\}$, has $N_r=10+2k$ and $N_\ell=14+2k$, so that Lev3 uses $N_r=16$ and $N_\ell=20$.

\begin{figure*}[!t]
 \centering
 \includegraphics[width=\textwidth]{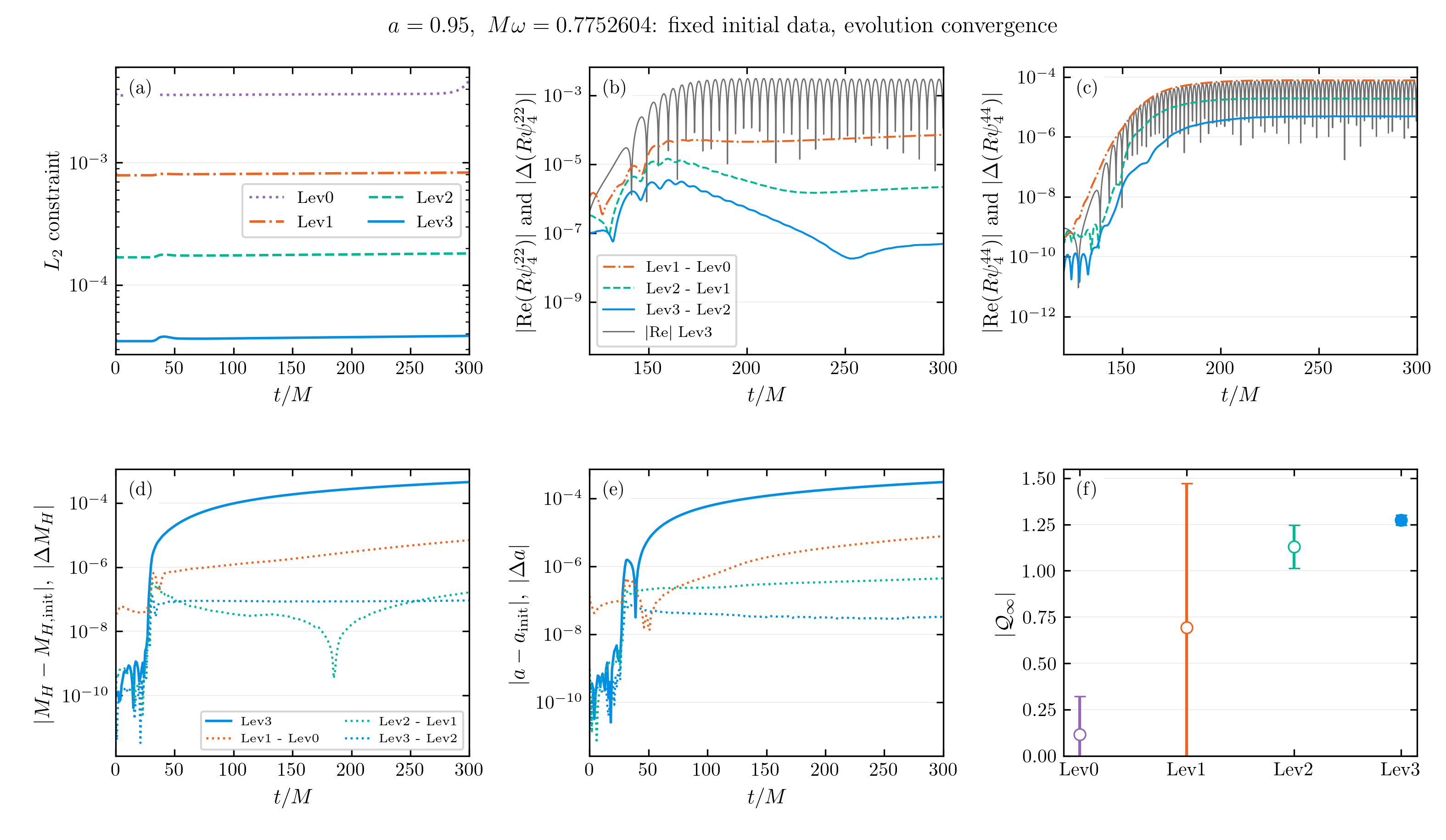}
 \caption{Fixed-initial-data convergence study at $a=0.95$, $M\omega=0.7752604$. Upper row: generalized-harmonic $L_2$ constraint, averaged in $0.1M$ bins; absolute real parts of the Lev3 parent and daughter waveforms at $R=93M$, with absolute complex differences between adjacent resolutions. Lower row: changes in horizon mass and spin at Lev3, differences between adjacent resolutions, and extrapolated $|\mathcal Q_\infty|$. Open markers fail the extrapolation criteria while the filled Lev3 marker passes. Error bars describe fit and extrapolation sensitivity.}
 \label{fig:NRconvergence}
\end{figure*}

Figure~\ref{fig:NRconvergence} compares constraint violations, waveforms, horizon diagnostics, and the extrapolated coupling. The values of $|\mathcal Q_\infty|$ are $0.11765$, $0.69430$, $1.13040$, and $1.27324$ at Levs 0--3. Lev0 and Lev1 have strong extrapolation radial-order dependence, while Lev2 has a $10.26\%$ spread when omitting a single radius at a time from the extrapolation procedure. These three levels fail the extrapolation criteria defined below and are shown as open markers. Lev3 passes with a maximum tested complex-intercept (defined below) variation of $2.10\%$. The Lev2--Lev3 difference is nevertheless $12.96\%$ in the complex coupling and $11.22\%$ in magnitude.

For this case, the nonlinear initial-data solver uses absolute and relative tolerances $10^{-5}$ and $10^{-6}$, and linear-solver tolerances $10^{-7}$ and $10^{-6}$. The evolution uses an ordinary differential equation tolerance of $10^{-8}$ and outputs the waveform and horizon diagnostics every $0.1M$ and $M$, respectively. Horizon drift is measured from the initial Christodoulou mass and spin magnitude.

\subsection{Complex waveform fits}
At each extraction radius, we fit the waveforms over a quasi-stationary interval using
\begin{subequations}
\label{eq:fitform}
\begin{align}
 r\psi_4^{22}(t)
 &=\mathcal A^{\psi_4}_{22}e^{-i\omega t}
  +\mathcal B^{\psi_4}_{22}e^{i\omega t}
  +\mathcal C^{\psi_4}_{22},\\
 r\psi_4^{44}(t)
 &=\mathcal A^{\psi_4}_{44}e^{-2i\omega t}
  +\mathcal B^{\psi_4}_{44}e^{2i\omega t}
  +\mathcal C^{\psi_4}_{44}+Dx \nonumber\\
 &\quad+\sum_{\sigma=\pm1}\sum_{k=0}^{d}
          b_{\sigma k}x^k e^{i\sigma\omega t},\\
 x&=(t/M-250)/100,
\end{align}
\end{subequations}
where all coefficients are complex. The parent uses trapezoidally time-weighted least squares. The daughter uses unweighted complex least squares with scaled design columns. Fits must have full rank and a scaled condition number below $10^8$. The terms at both signs of the carrier frequency and the constant and slowly varying terms separate the $2\omega$ signal from other waveform content.

\begin{figure*}[!t]
 \centering
 \includegraphics[width=\textwidth]{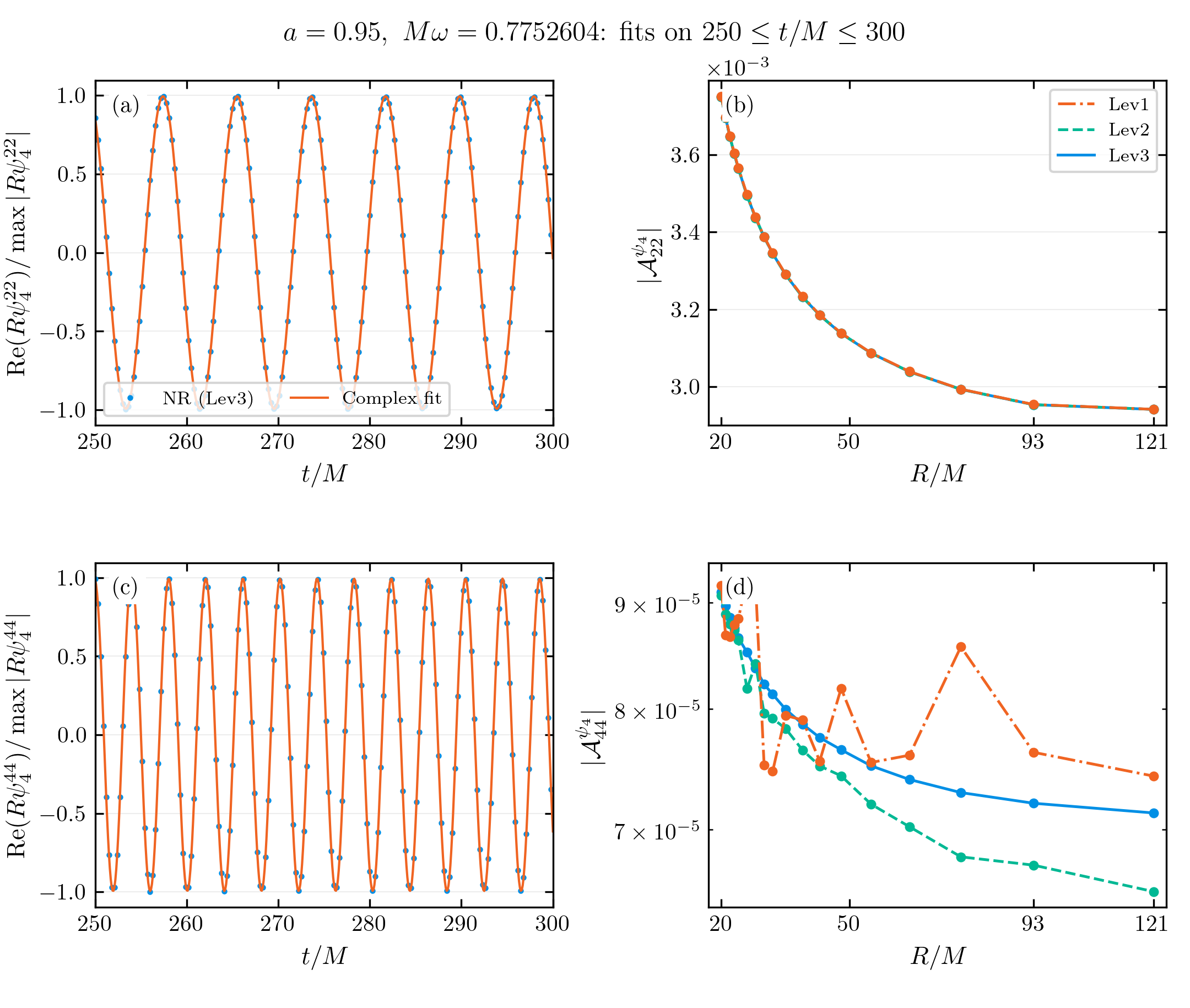}
 \caption{Complex fits and extraction-radius dependence for the cases shown in Fig.~\ref{fig:NRconvergence}. Left: real parts of the Lev3 parent and daughter at $R=93M$, each normalized by its own maximum complex magnitude, with fits over $250\leq t/M\leq300$. Right: fitted magnitudes at Levs 1--3 versus extraction radius. The under-resolved Lev0 is omitted here. Only radii through $121M$, which pass the fit-stability criteria, are displayed. The finite-radius fits shown here use a common coordinate-time interval, unlike the common-retarded-time extrapolation used for $\mathcal Q_\infty$.}
 \label{fig:NRfits}
\end{figure*}

For the illustration in Fig.~\ref{fig:NRfits}, the reference interval is $250\leq t/M\leq300$ with $d=3$. At Lev3 and $R=93M$, the normalized parent and daughter residuals are $0.280\%$ and $0.357\%$. Between $R=93M$ and $121M$, the fitted coupling magnitude changes by $-0.244\%$ and its phase changes by $-1.44$ degrees, giving a $2.53\%$ complex difference.

\subsection{Extrapolation to infinity and sample selection}
\label{app:extrapolation}
The extrapolation of the finite-radius $\mathcal{Q}(R)$ proceeds in four steps: (i) fit the parent and daughter amplitudes at five extraction radii over a common retarded-time interval, (ii) compute $\mathcal Q(R)$ at each radius, (iii) extrapolate $\mathcal Q(R)$ to $R\to\infty$, and (iv) repeat (i)--(iii) over a fixed set of fitting variants to obtain a sensitivity envelope and to decide whether the case is sufficiently robust to be retained.

Waveforms are stored at coordinate radii $R/M\in\{55,64,76,93,121\}$. The finite wave train reaches each radius at a different coordinate time, so fitting all radii over one interval of $t$ would compare different portions of the train. We therefore fit over a common interval of the approximate retarded time $u=t-D(R)$, with $D$ the Boyer--Lindquist tortoise coordinate $r_*$ of the background. In units $M=1$,
\begin{equation}
\begin{aligned}
 r_\pm&=1\pm\sqrt{1-a^2},\\
 D(R)&=R+\frac{2r_+}{r_+-r_-}\ln\!\left(\frac{R-r_+}{2}\right)\\
      &\quad\quad\,\,-\frac{2r_-}{r_+-r_-}\ln\!\left(\frac{R-r_-}{2}\right),
\end{aligned}
\label{eq:null_delay}
\end{equation}
which reduces to $D(R)=R+2\ln(R/2-1)$ for $a=0$. At each radius the parent and daughter are fitted with Eq.~\eqref{eq:fitform} over the reference interval $110\leq u/M\leq160$.

The five values $\mathcal Q(R)$ are then fitted by complex least squares to
\begin{equation}
 \mathcal Q(R)=\mathcal Q_\infty+c_1\,(100M/R)+c_2\,(100M/R)^2,
 \label{eq:radial_extrapolation}
\end{equation}
where the factor $100M$ is introduced to condition the design matrix. The value we report for each run is the intercept $\mathcal Q_\infty$ of this fit when the waveforms are fitted over $110\leq u/M\leq160$ with the cubic daughter envelope [$d=3$ in Eq.~\eqref{eq:fitform}] and all five radii are used.

The reported $\mathcal Q_\infty$ depends on three choices: the retarded-time interval over which the waveforms are fitted, the polynomial degree $d$ of the daughter envelope in Eq.~\eqref{eq:fitform}, and the set of radii used in Eq.~\eqref{eq:radial_extrapolation}. To estimate how much each choice matters, we recompute the intercept with one choice altered at a time, which gives fifteen alternative intercepts. Altering the interval and envelope gives nine: the four intervals $u/M\in(105,155)$, $(105,160)$, $(110,155)$, $(100,160)$ with $d=3$, and all five intervals with $d=2$. Altering the radial fit gives six: a linear fit in $1/R$ using all five radii, and quadratic fits that each omit one of the five radii. Writing $\mathcal Q_v$ for these alternative intercepts, the amplitude bar on a point in Figs.~\ref{fig:NRcoupling} and~\ref{fig:qnm_coupling_comparison} extends from the smallest to the largest value of $|\mathcal Q_v|-|\mathcal Q_\infty|$, and the phase bar from the smallest to the largest value of $\arg(\mathcal Q_v/\mathcal Q_\infty)$.

A value of the coupling is accepted when it passes the following checks: (i) the spread of the alternative intercepts $\mathcal Q_v$ described above must be below $10\%$ of $|\mathcal Q_\infty|$, (ii) the largest residual of the fit in Eq.~\eqref{eq:radial_extrapolation} must be below $5\%$, (iii) the residuals of the parent and daughter waveform fits must each be below $10\%$, (iv) a daughter fit from which the central $10M$ of the interval is withheld must reproduce the withheld data to within $20\%$, and (v) all fits must be of full rank with a scaled condition number below $10^8$. 

Twenty-one otherwise eligible runs fail these checks when the common interval $110\leq u/M\leq160$ is used. Most are at the lowest frequency, $M|\omega|=0.18$, where the wave train is short, or at the highest prograde frequencies for $a=0.95$, where the daughter settles late. For these runs we choose the fitting interval instead from a fixed grid, with lower endpoint between $80M$ and $155M$ and upper endpoint $160M$ or $170M$, selecting the interval that minimizes the largest of the diagnostics listed above. The fifteen alternative intercepts are then rebuilt around the selected interval. The $a=0.95$ prograde runs, for example, select $140\leq u/M\leq170$. Seventeen of the twenty-one runs pass every check with their selected interval. The remaining four at $(a,M\omega)=(0,0.76)$, $(0.25,0.76)$, $(0.9,0.18)$, and $(0.95,0.85)$ still exceed at least one threshold.

\section{A scalar model for qualitative interpretation}\label{app:toymodel}
We describe the construction and numerical solution of the one-dimensional scalar model scattering problem discussed briefly in the main text. 

\subsection{Radial potential}
We consider massless scalar waves of the form
\begin{align}
    \Phi(t,r,\phi)=e^{-i\omega t +im\phi}Y(r),
\end{align}
where, for simplicity, we neglect any $\theta$ dependence in $\Phi$ and the radial wavefunction $Y(r)$ is governed by the equation
\begin{align}
    \frac{\ed^{2}Y_{\ell m}}{\ed r_{*}^2}-U(r;\omega,\ell,m)Y_{\ell m}=F(r).\label{eq:toyradial}
\end{align}
The radial potential $U(r;\omega,\ell,m)$ is constructed phenomenologically from the propagation of massless scalar plane waves on the Kerr background in the high-frequency Wentzel--Kramers--Brillouin (WKB) limit.

We consider the WKB ansatz $\Psi(x)=A(x)e^{iS(x)/\epsilon}$, where $A(x)$ is a slowly varying amplitude, $S(x)$ is a rapidly varying phase, and $\epsilon\ll1$ is a dimensionless bookkeeping parameter. To leading order $\mathcal{O}(\epsilon^{-2})$, the wave equation $\Box\Psi=0$ reduces to $k_{a}k^{a}=0$, where $k=\ed S$ is a $1$-form normal to hypersurfaces of constant phase $S(x)$. In the high-frequency WKB limit, the level sets of $S(x)$ are null hypersurfaces whose generators are null geodesics. The equation of motion $k^{a}k_{a}=0$ is a Hamilton--Jacobi (HJ) equation for the phase $S$. The Killing symmetries of Kerr spacetime result in the HJ equation being separable~\cite{Carter:1968rr}. In particular, the ansatz
\begin{align}
    S(x) = -\omega t+m\phi+S_{r}(r)+S_{\theta}(\theta)
\end{align}
leads to the separated radial equation~\cite{Carter:1968rr,Chandrasekhar:1985kt}
\begin{align}
    \pa{\frac{\ed S_{r}}{\ed r_{*}}}^2&=\br{\omega-\frac{am}{r^2+a^2}}^2-\frac{\Delta K}{\pa{r^2+a^2}^2}\label{eq:radialHJ}
\end{align}
where $\Delta=r^2-2r+a^2$, $\mathrm dr_*/\mathrm dr=(r^2+a^2)/\Delta$, and $K$ is the Carter separation constant. For geodesic motion in the Kerr background, $K$ is generated by a rank-two Killing symmetry~\cite{Carter:1968rr,Walker:1970un}. In our phenomenological model, we instead take $K\to \Lambda_{\ell}=\ell(\ell+1)$ to introduce dependence on the angular quantum number $\ell$.

The WKB ansatz $Y(r_{*})=A(r_{*})e^{iS_{r}(r_{*})/\epsilon}$ substituted into the homogeneous radial wave equation
\begin{align}
    \frac{\ed^2 Y}{\ed r_{*}^2}-U(r_{*})Y=0,\label{eq:toyradialhom}
\end{align}
reduces to $S_{r}'(r_{*})^2=-\bar{U}(r_{*})$, assuming $U=\epsilon^{-2}\bar{U}$. Matching the high-frequency WKB limit of Eq.~\eqref{eq:toyradialhom}  to Eq.~\eqref{eq:radialHJ} then motivates the choice of potential
\begin{align}
    U(r;\omega,\ell,m)&=\frac{\Delta \Lambda_{\ell}}{\left(r^2+a^2\right)^2}-\left[\omega-\frac{a m}{r^{2}+a^{2}}\right]^2.\label{eq:toypotential}
\end{align}

\subsection{Quadratic source and daughter solution}
The in and up homogeneous solutions to the scalar model obey
\begin{subequations}
    \begin{gather}
        Y^{\rm{in}}
        \sim e^{-ik_{H}r_{*}},\,{r_{*}\to-\infty},\\
        Y^{\rm{up}}
        \sim e^{i\omega{r}_{*}},\,r_{*}\to\infty,
    \end{gather}
\end{subequations}
where $k_H=\omega-m\Omega_H$ and $\Omega_H=a/(r_+^2+a^2)$.
We solve for the in- and up-mode solutions with $(2,2,\omega)$ and $(4,4,2\omega)$ via a fourth-order Runge--Kutta numerical integration. The Wronskian $W$ is constructed from the $(4,4,2\omega)$ homogeneous solutions as
\begin{align}
    W=Y^{\rm{in}}_{44}\partial_{r_{*}}Y^{\rm{up}}_{44}-Y^{\rm{up}}_{44}\partial_{r_{*}}Y^{\rm{in}}_{44}.
\end{align}
The daughter $(4,4,2\omega)$ particular solution is given by convolution of the source $F(r)\equiv F_{44}$ with the $(4,4,2\omega)$ Green's function,
\begin{align}
    Y^{(2)}_{44}(r_{*})&=\frac{1}{W}\left[Y_{44}^{\rm{up}}(r_{*})\int_{-\infty}^{r_{*}}Y_{44}^{\rm{in}}(\xi)F_{44}(\xi)\ed\xi\right.\nonumber\\
    &\qquad\left.+ Y_{44}^{\rm{in}}(r_{*})\int_{r_{*}}^{\infty}Y_{44}^{\rm{up}}(\xi)F_{44}(\xi)\ed\xi\right].\label{eq:sourcedtoysol}
\end{align}
We evaluate these integrals using a cumulative Simpson quadrature. The source is constructed from the parent in-mode as
\begin{align}
    F_{44}=V_{2}V_{4}\pa{f_{r}F_{r}+f_{t}F_{t}+f_{\times}F_{\times}+f_{V}F_{V}},\label{eq:toysource}
\end{align}
where
\begin{equation}
\begin{gathered}
    F_{r}=(D_{r}Y)^2,
    F_{t}=(D_{t}Y)^2,F_{V}=V_{2}Y_{22}^2,\\
    V_{\ell}=\frac{\Delta\Lambda_{\ell}}{\pa{r^2+a^2}^2},
    D_{t}Y=\left(\omega-\frac{am}{r^2+a^2}\right)Y_{22},\label{eq:DtY}\\
    D_{r}Y=\partial_{r_{*}}Y_{22},F_{\times}=-2iD_{t}YD_{r}Y,
\end{gathered}
\end{equation}
with $m=2$ and $Y=Y_{22}^{\rm{in}}$. The $F_{i}$ terms in Eq.~\eqref{eq:toysource} are products of $Y_{22}$ and its time and radial derivatives. The overall factor of $V_{2}V_{4}$ acts as a volume-form-like wave compression factor in the strong field. Furthermore, the $V_{2}V_{4}$ factor ensures the source is regular on the outer horizon $r_{+}$ and decays as $F_{44}=\mathcal{O}(r^{-4})$ as $r\to\infty$. The $f_{i}$ are real constants determined by globally fitting to the coupling coefficient data from the NR simulations.

\subsection{Numerical implementation}

\begin{figure}[hbt!]
    \centering
    \includegraphics[width=1.0\columnwidth]{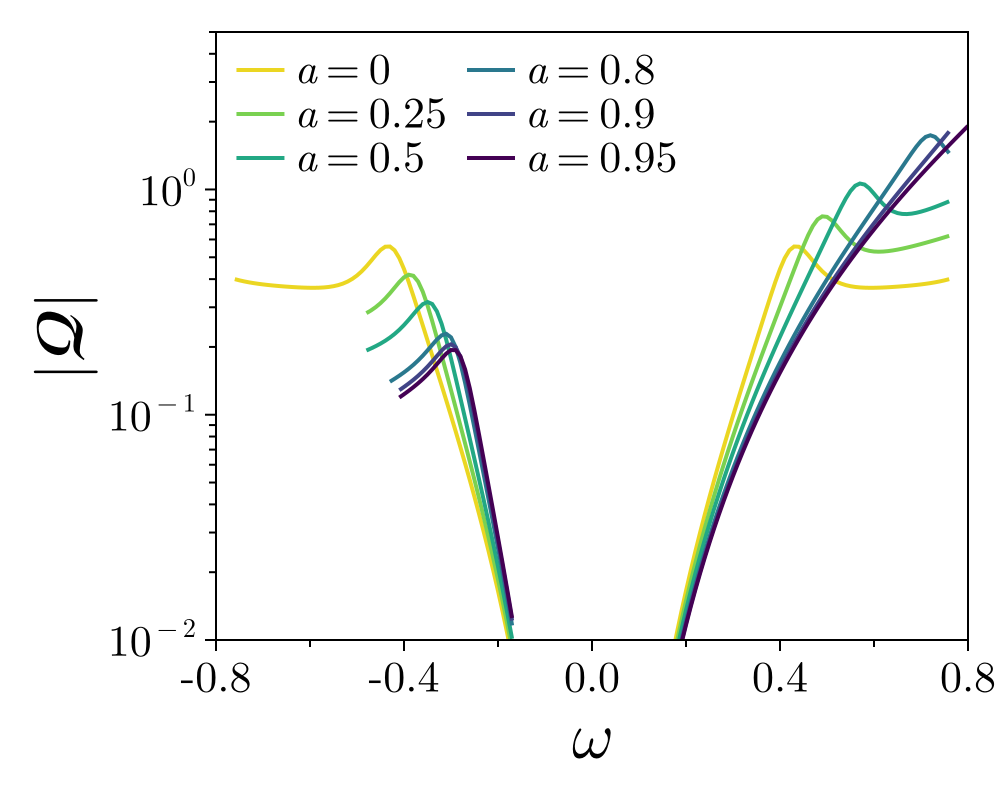}
    \caption{Scalar model quadratic coupling  \eqref{eq:toyQ} as a function of angular frequency $\omega$ for several black-hole spins $a$ (see legend).}
    \label{fig:toycoupling}
\end{figure}

To obtain the in- and up-modes, we numerically integrate the homogeneous scalar wave equation as a first-order system uniformly in the tortoise coordinate $r_{*}$ using a fourth-order Runge--Kutta (RK$4$) scheme. Since $r_{*}(r_{+})=-\infty$, the inner boundary of the domain is taken to be just outside $r_{+}$, so that the integration domain is $r\in[r_{\min},r_{\max}]$, where $r_{\min}=r_{+}+\delta$, $\delta\ll M$, and $r_{\max}\gg M$. The numerical integration of the in (up) mode starts at $r=r_{\min}$ ($r=r_{\max}$) with unit normalization $Y=1$ at the starting radius. The underlying solutions used to make Fig.~\ref{fig:toycoupling} and Fig.~\ref{fig:toypotential} were obtained with $\delta=10^{-10}M$, $r_{\max}=4000M$, and uniform tortoise coordinate step $\Delta{r_{*}}\equiv h=0.025M$. Convergence with respect to $\delta$, $h$, and $r_{\max}$ is demonstrated in the next subsection.

The scalar model quadratic coupling coefficient is computed by fitting the homogeneous $(2,2,\omega)$ in-mode solution $Y^{(1)}_{22}$ and the $(4,4,2\omega)$ particular solution $Y^{(2)}_{44}$ at large radii $r\sim r_{\max}\gg M$ over the domain $[0.95\,r_{\rm{max}},r_{\rm{max}}]$. We fit to the ans\"atze
\begin{subequations}
    \begin{align}
        Y^{(1)}_{22}=A^{(1)}_{\rm{in}}e^{-i\omega r_{*}}+A_{\rm{out}}^{(1)}e^{i\omega r_{*}}+C_{1},\\
        Y^{(2)}_{44}=A^{(2)}_{\rm{in}}e^{-2i\omega r_{*}}+A_{\rm{out}}^{(2)}e^{2i\omega r_{*}}+C_{2},
    \end{align}\label{eq:toyansatze}
\end{subequations}
where the $A_{\rm{in}}^{(2)}$ and $C_{i}$ terms are included to account for numerical contamination giving rise to unphysical constant terms and an unphysical incoming component in the second-order wavefunction. The scalar-model quadratic coupling is defined as
\begin{align}
     \mathcal{Q}(\omega)=\frac{A_{\rm out}^{(2)}(2\omega)}{\br{A_{\rm out}^{(1)}(\omega)}^{2}}.\label{eq:toyQ}
\end{align}

The coefficients $f_{i}$ in the source term Eq.~\eqref{eq:toysource} are determined by uniformly weighted linear least-squares fits of the scalar-model quadratic coupling $\mathcal{Q}$ to the values obtained from NR. We exclude near-resonant couplings satisfying $0.8\leq M\ab{\omega}/\br{\mathrm{Re}(\omega_{4,\pm4,0})/2}\leq1.3$ from the scalar model fit, leaving a total of $161$ data points used in the fit over spins $a\in\{0,0.25,0.5,0.55,\dots,0.95\}$ and frequencies $0.177\leq M\ab{\omega}\leq0.764$. The result is
\begin{equation}
\begin{gathered}
    f_r=3.48,\,
    f_t=2.97,\\
    f_\times=-1.44,\,
    f_V=-5.04.
\end{gathered}\label{eq:toycoeffs}
\end{equation}

We plot in Fig.~\ref{fig:toycoupling} the quadratic coupling spectrum for the same NR spin cases as in Fig.~\ref{fig:NRcoupling} up to the maximum value of $\ab{\omega}$ along each curve at fixed spin on a fixed branch.

\subsection{Wave reflection, transmission, and suppression in the scalar model}

\begin{figure*}[hbt!]
    \centering
    \includegraphics[width=\textwidth]{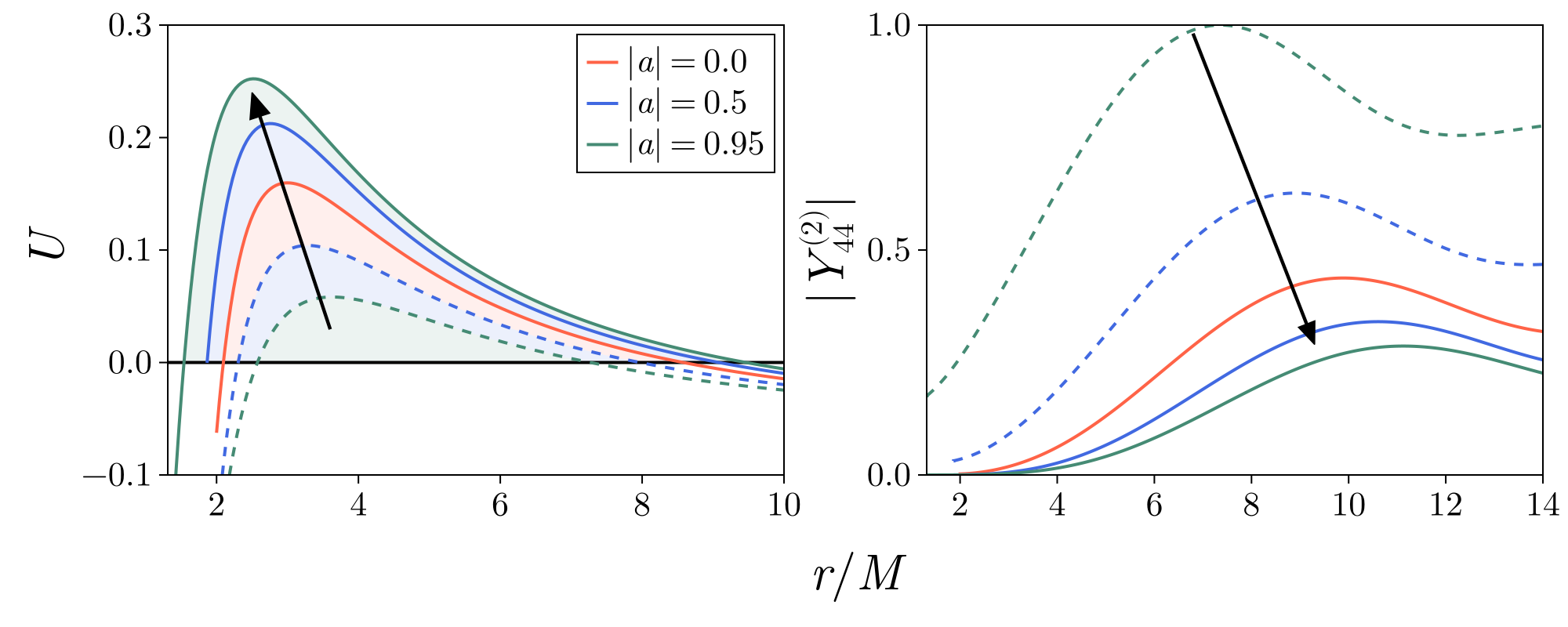}
    \caption{Left: scalar model potential \eqref{eq:toypotential} with $\ell=m=2$ and $M\ab{\omega}=0.25$ for several values of the black-hole spin $a$ (see legend). The color of each curve encodes the value of $\ab{a}$ and the line style indicates either prograde (solid) or retrograde (dashed) wave motion. The black horizontal curve corresponds to $U=0$; all other curves start at the outer horizon. The shaded regions are classically forbidden. Right: corresponding second-order wavefunctions sourced by the in-mode homogeneous solution (normalized by the maximum value of the $a=-0.95$ wavefunction). As the spin of the black hole increases from $-0.95$ to $0.95$ (indicated by the black arrows), both the classical outer turning radius and the radial extent of the forbidden region increase, leading to wave scattering further outside and increased wavefunction suppression inside the strong-field region.}
    \label{fig:toypotential}
\end{figure*}

In Fig.~\ref{fig:toypotential}, we hold $\omega>0$ and vary signed spin, with $a>0$ and $a<0$ denoting prograde and retrograde motion, respectively. The classical turning radii $R_\pm$ satisfy $U(R_\pm)=0$. An incident wave from past null infinity is partially reflected at the outer turning radius and partially transmitted into the classically forbidden region $R_-<r<R_+$ in which $U>0$. In this region, the WKB wave amplitude is exponentially attenuated before reaching the horizon.

As illustrated in the left panel of Fig.~\ref{fig:toypotential}, increasing signed spin at low frequencies moves the outer turning point further outwards and increases the barrier height and radial extent, each of which reduces the parent amplitude reaching the strong-field region. The daughter profiles in Fig.~\ref{fig:toypotential} are correspondingly weaker at small radius, with their maxima shifted outwards. Opposite effects occur for increasing the magnitude of prograde and retrograde spin, leading to the diminished (enhanced) coupling with spin on the prograde (retrograde) branch in the scalar model. At sufficiently high frequencies, the forbidden region disappears and this turning-point interpretation no longer applies. 

\subsection{Convergence tests}\label{app:toyconvergence}
\begin{figure}[hbt!]
    \centering
    \includegraphics[width=1.0\columnwidth]{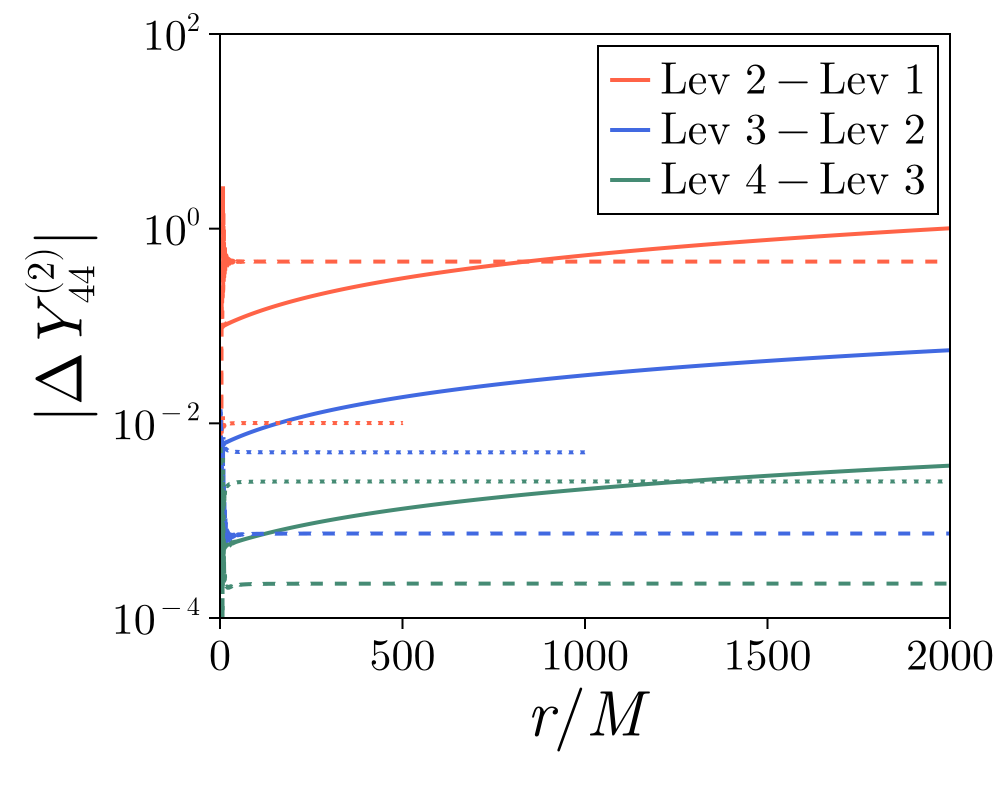}
    \caption{Scalar model convergence in the second-order wavefunction for $a=0.95$ and $M\omega=1$ as one of $h$, $\delta$, and $r_{\max}$ is refined. Each curve corresponds to the pointwise difference between $Y_{44}^{(2)}(r)$ computed on two adjacent refinement levels divided by the higher-resolution wavefunction, with level $4$ the most refined. The three families of curves (solid, dashed, dotted) exhibit convergence. The solid curves have $\delta=10^{-10}M$, $r_{\max}=2000M$, and $h/M\in\{0.2,0.1,0.05, 0.025\}$. The dashed curves have $h=0.025M$, $r_{\max}=4000M$, and $\delta/M\in\{10^{-1},10^{-4},10^{-7},10^{-10}\}$. The dotted curves have $h=0.025M$, $\delta=10^{-10}M$, $r_{\max}/M\in\{500,1000,2000, 4000\}$. The dotted curves have a radial extent given by the lesser value of $r_{\max}$ between the two adjacent levels. Convergence in the corresponding values of $\mathcal{Q}$ is demonstrated in Table~\ref{tab:toyconvergence}.}
    \label{fig:toyconvergence}
\end{figure}

We demonstrate convergence of the second-order wavefunction $Y_{44}^{(2)}$ and the quadratic coupling $\mathcal{Q}$ for $a=0.95$ and $M\omega=1$. We plot in Fig.~\ref{fig:toyconvergence} differences in the second-order wavefunction as one of $h$, $\delta$, and $r_{\max}$ is refined, with the other two parameters held fixed. We list in Table~\ref{tab:toyconvergence} the corresponding differences in $\mathcal{Q}$. As each numerical parameter is refined, both the wavefunction and quadratic coupling exhibit convergence.
\begin{table}[hbt!]
\caption{\label{tab:toyconvergence}
Convergence of the scalar model quadratic coupling for $a=0.95$ and $M\omega=1$ as one of $h$, $\delta$, and $r_{\max}$ is refined with the others held fixed. See the caption of Fig.~\ref{fig:toyconvergence} for the values of the numerical parameters in each convergence test. The header of each column denotes which numerical parameter is refined, and each entry corresponds to the percentage difference in the quadratic coupling between two adjacent refinement levels. The values quoted correspond to the nine different curves in Fig.~\ref{fig:toyconvergence}, with $\Delta\mathcal{Q}_{12}\equiv 1-\mathcal{Q}_{\rm{Lev}1}/\mathcal{Q}_{\rm{Lev}2}$ and likewise for $\Delta\mathcal{Q}_{23}$ and $\Delta\mathcal{Q}_{34}$.}
\begin{ruledtabular}
\begin{tabular}{cccc}
 &$h$ &$\delta$ &$r_{\max}$\\
\hline
$\ab{\Delta\mathcal{Q}_{12}}$& $7.2\times10^{-1}$ & $1.3\times10^{-1}$ & $2.0\times10^{-4}$\\
$\ab{\Delta\mathcal{Q}_{23}}$& $1.7\times10^{-2}$ & $1.0\times10^{-4}$ & $2.1\times10^{-5}$\\
$\ab{\Delta\mathcal{Q}_{34}}$& $5.4\times10^{-4}$ & $3.7\times10^{-7}$ & $1.8\times10^{-5}$
\end{tabular}
\end{ruledtabular}
\end{table}

\end{document}